# The Shape of Reactive Coordination Tasks

Ido Ben-Zvi
Technion
Haifa, Israel
iddobz@gmail.com

Yoram Moses
Technion
Haifa, Israel
moses@ee.technion.ac.il

## ABSTRACT
This paper studies the interaction between knowledge, time and coordination in systems in which timing information is available. Necessary conditions are given for the causal structure in coordination problems consisting of orchestrating a set of actions in a manner that satisfies a variety of temporal ordering assumptions. Results are obtained in two main steps: A specification of coordination is shown to require epistemic properties, and the causal structure required to obtain these properties is characterised via "*knowledge gain*" theorems. A new causal structure called a *centibroom* structure is presented, generalising previous causal structures for this model. It is shown to capture coordination tasks in which a sequence of clusters of events is performed in linear order, while within each cluster all actions must take place simultaneously. This form of coordination is shown to require the agents to gain a nested common knowledge of particular facts, which in turn requires a centibroom. Altogether, the results presented provide a broad view of the causal shape underlying partially ordered coordinated actions. This, in turn, provides insight into and can enable the design of efficient solutions to the coordination tasks in question.

## Categories and Subject Descriptors
[**Artificial intelligence**]: Knowledge representation and reasoning — *Reasoning about belief and knowledge, Causal reasoning and diagnostics*; [**Artificial intelligence**]: Distributed artificial intelligence — *Cooperation and coordination, multi-agent systems*; [**Distributed computing methodologies**]

## 1. INTRODUCTION
Coordinated action in distributed and multi-agent systems is closely related to knowledge and epistemic states. As a particular example, linearly ordered actions require nested knowledge. Namely, suppose that the occurrence of event $e$ is guaranteed to trigger a response by each of the agents 1, 2, and 3, and, moreover, they must act in this order: first 1, then 2, and finally 3. Then, in a precise sense, $K_3 K_2 K_1 \mathsf{occ}(e)$ (which we read as "*agent 3 knows that 2 knows that 1 knows that e has occurred*") must hold when agent 3 acts [3]. This generalises from three agents to any finite number. In the theory of distributed systems, asynchronous systems, in which agents have no clock and no timing information is available, receive a great deal of attention [1, 17]. In such systems, Chandy and Misra's celebrated *Knowledge Gain* theorem [7] captures the necessary condition for attaining nested knowledge of this form. Roughly speaking, it implies the following. Suppose that a spontaneous event $e$ takes place at agent 0's site in an asynchronous system. Then $K_3 K_2 K_1 \mathsf{occ}(e)$ can hold only after a message chain is formed, that starts from agent 0 after $e$ occurs, and passes through 1 and then through 2 to agent 3. (The message chain may pass through other sites as well; but it must visit these agents in the specified order.) As a result, the only way to coordinate a linearly ordered response to the event $e$ in an asynchronous system is via such a message chain. This theorem captures the shape of the causal structure that underlies linear coordination.

The presence of clocks and timing information can greatly facilitate coordination tasks in distributed and multi-agent systems. In [3, 4] we initiated a study of coordination in a *synchronous* model, where agents have access to a global clock, and, for each particular channel, there is an upper bound on the time messages can spend in transit. In the presence of clocks the passage of time can be used to derive information about events at remote sites. As a result, message chains are not the only way to attain nested knowledge. A knowledge gain theorem capturing subtle interplay of communicated messages, the guaranteed bounds, and the passage of time is given in [3]. It shows that the causal "shape" underlying nested knowledge is captured by a structure called a *centipede* (see Figure 2). In a precise sense, this is the "synchronous" analogue of a message chain.

The connection between coordination and epistemics manifests itself beyond the connection between linearly ordered actions and nested knowledge. Halpern and Moses showed that simultaneous actions are very closely related to common knowledge [14]: If a set of agents B can be guaranteed to all perform a particular action a *simultaneously* whenever any of them perform it, then when they perform a they have *common knowledge* that it is being performed. If the action a is performed only in response to a particular spontaneous event $e$, then they must also have common knowledge that $e$ has occurred. Using more formal notation, if we denote common knowledge to B by $C_B$, then attaining $C_B\mathsf{occ}(e)$ is a prerequisite for performing a. While common knowledge (as well as simultaneity) cannot be attained in asynchronous settings [7, 14], it can often be attained in the presence of clocks and time guarantees. A causal structure called a *broom* (Figure 3) was shown to be necessary for gaining common knowledge in systems with clocks [3]. Both centipedes and brooms are defined in terms of two relations (the different styled edges in Figures 2 and 3): *syncausality*, which captures message chains in the synchronous model, and *bound guarantee*, which provides the means to account for the information obtained by the passage of time. (We review the definitions of these relations and structures in Section 2.2.)

Characterizing the causal shape underlying coordination tasks provides insight into the structure of their solutions, and often enables the design of optimal solutions for such problems. In Section 3 we demonstrate this with a novel application by deriving an optimal solution to the *distributed snapshot* problem of [6] in the synchronous setting, based on the connection between brooms and simultaneous response.

In this paper, we extend the analysis of coordination in the synchronous model, to handle much more general forms of coordination. We follow the same scheme as above: Relate a class of coordination problems to a set of corresponding epistemic states, and then study the causal structure underlying these epistemic states via proving knowledge gain theorems.

Consider the following example, which constitutes a variation on one discussed by Chwe [8] and is related to numerous studies of information flow and agency in social networks [13, 22, 19].

EXAMPLE 1. *Under the yoke of the Roman conquerer, the repressed Judean people are bitter and rebellious. As a population, the agents are partitioned into several groups by their tendency to revolt: there is an instigator, and there are the hardline ideologists, the unsatisfied crowds, and the supporters of the old regime.*

- *The **instigator** is highly unpredictable. It may start a revolt at any time, independent of any other event.*

- *A **hard liner** will revolt if it knows that the instigator and all of the other hard liners are revolting together.*[1]

- *A member of the **unsatisfied masses** will revolt if it knows that the instigator, the hard liners and all of the other members of the unsatisfied masses are revolting.*

- *A supporter of the **old regime** will revolt only when it knows that all other members of the population are revolting.*

*By means of sun clocks and camel-borne messages, the agents form a synchronous system with upper limits on message transmission times. If the agents are continually communicating with each other, then it is possible to arrange for a rebellion to start a finite number of days after the instigator revolts.*[2]

*The question we ask is — what pattern of communication would suffice to ensure that the whole population revolts, while keeping communication to a minimum, in the sense that unneeded messages are not sent (so as not to arouse the suspicion of the infamous Roman crucifixion police)?*

The problem faced by the Judeans in Example 1 transcends both of the coordination tasks discussed earlier, because it involves both a linear ordering *and* simultaneity of events. More precisely, it involves a sequential ordering of clusters of responses, where the actions in every cluster are performed simultaneously. We will define a corresponding class of coordination problems, called *Ordered Joint Response* (OJR). In general, the sets of acting agents in every cluster will not be assumed to be disjoint. We will show that solving OJR requires attaining nested common knowledge of the form

$$C_{B^k} C_{B^{k-1}} \cdots C_{B^1} \mathsf{occ}(e).$$

The main technical contribution of the paper is a nested common knowledge gain theorem (NCKG) for the synchronous model. It captures the causal structure underlying NCKG by a new form called a *centibroom* (see Figure 5), which is a hybrid structure, combining the centipede with brooms. Since the centibroom is necessary for getting the Judean groups from Example 1 to revolt in the proper order, it is also the minimal communication pattern.

Once we have established how linear sequences of joint responses can be ordered, we will take a bigger step and consider the general problem of ordering events according to any pre-specified ordering. Consider the following update on the situation in Judea.

EXAMPLE 2. *The situation is just as dire as in Example 1, but the social standing is more complex, as there are now three instigators, and the hardline ideologists are divided among themselves into two opposing groups: the Peoples Front of Judea (PFJ) and the Judean People's Front (JPF).*

- *The three instigators are: Jedediah, Jeremiah, and Brian - each of them operating on its own as before.*

- *A member of the PFJ will revolt if it knows that Jedediah, Jeremiah and the rest of the PFJ's members are revolting.*

---

[1] Assume for each of the groups in the population that common knowledge of "stalemate" is resolved by revolting. I.e., if it is common knowledge among the hardliners that the instigator is revolting, then each hardliner revolts too.

[2] In fact, the proposed solution to the distributed snapshot problem, discussed in Section 3, coud be used to achieve this in minimal time.

- *A member of the JPF will revolt if it knows that Brian and the rest of the JPF's members are revolting.*

- *The unsatisfied masses and the supporters of the old regime act as before.*

*Figure 1 sums up the revolt dependencies as a directed acyclic graph. Once again we ask what pattern of communication would push the population into rebellion (provided that enough instigators revolt) while keeping communication to a minimum.*

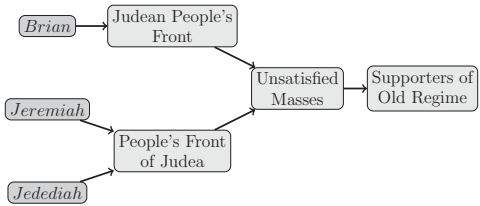

Figure 1: Judea, 71AD

In Example 2, if both Jedediah and Brian revolt (but not Jeremiah), then the country will not be swept by rebellion. The members of the JPF will also revolt, as they only look to Brian for guidance. But those of the PFJ will abstain — waiting for Jeremiah to revolt as well before joining in. The unsatisfied masses will see that the PFJ is not joining in, and will prefer to stay at home. In contrast, if all three instigators revolt (and there is sufficient communication to spread the word) then both hardliner factions will revolt too — eventually leading the unsatisfied masses to revolt, and even the supporters of the old regime to follow in their wake. Note that even though the members of the PFG all revolt simultaneously, as do all members of the JPF, these two simultaneous joint responses need not occur at the same time.

In order to derive the necessary epistemic state and communication pattern for solving such problems, we will consider *General Ordered Response* problems (GOR), in which a weak ordering among responses is specified by a general directed graph. An edge among two responses $\alpha$ and $\alpha'$ should imply that if $\alpha$ occurs at time $t$ in a given execution and $\alpha'$ occurs at $t'$, then $t \leq t'$. Note that we can encode both simultaneous sets of events, as well as an ordering on these sets, using the same partial order. We do this by making use of cycles on the graph, as all nodes on a cycle must be performed simultaneously. Thus, the GOR coordination problem can be used to specify a partial order on simultaneous clusters of responses. A we will show, while the OJR problem is solved by a new communication pattern called the centibroom, solutions to the more general GOR problem do not define a yet more complex structure that solves it. Rather, the communication pattern that characterizes it is best described as a set of unrelated centibrooms, thus capturing "the causal shape" of a very broad class of coordination problems in the synchronous model.

Our analysis is performed for *reactive* coordination tasks in which particular patterns of responses need to be performed in response to external triggering events. This is motivated by the fact that many distributed and multi-agent systems are embedded within a larger environment and need to coordinate their activities based on input that is supplied by it. This is true of an online bank, where customers may initiate transactions, a public safety application receiving the report of a smoke alarm activation, an online retailer (e.g. Amazon), a search engine (Google) that can accept requests, or a cloud computing application in which customers can submit computational tasks to be performed by the system. We focus on distributed settings in which activities in the system may be triggered by external events that are *spontaneous* as far as the system and its design are concerned. An external event of this sort may require a simple response by the system, but there are many cases in which it may trigger an extended transaction in which multiple events must take place, and these should be coordinated in various particular patterns.

This paper is organized as follows. The next section presents the model, and reviews the definitions of syncausality and bound guarantees, knowledge and common knowledge in distributed systems, and the centipede and broom structures from [3]. Section 3 illustrates the use of brooms by applying them to obtain an optimal distributed snapshot in the synchronous setting. Section 4 defines the ordered joint response problem, and relates the epistemic state of nested common knowledge to solutions to the problem. The notion of a centibroom is defined, and is used to capture nested common knowledge formulas, and sequential ordering of simultaneous responses. Finally, Section 5 defines the general ordered response problem, and states a theorem characterizing the shape of GOR solutions in terms of centibrooms. Section 6 closes the paper with discussion and further research.

## 2. BACKGROUND
## 2.1 Synchronous Networks

We focus on a simple synchronous setting in which agents are connected via a communication network and there are upper bounds on message transmission times. Agents share a global clock, and take steps at integer times. To analyze coordination in such a setting, we make use of the *interpreted systems* approach to modeling distributed systems (see [11]). Namely, we separate the definition of the environment for which protocols are designed, formally called the *context*, from the protocol being executed in that context. Formally, a context $\gamma$ is a tuple $(\mathcal{G}_0, P_e, \tau)$, where $\mathcal{G}_0$ is a set of initial global states, $P_e$ is a protocol for the environment, and $\tau$ is a *transition function*.[3] The environment is viewed as running a protocol (denoted by $P_e$) just like the agents; its protocol is used to capture nondeterministic aspects of the execution, such as the actual transmission times, external inputs into the system, etc. The transition function $\tau$ describes how the actions performed by the agents and by the environment change the global state.

A run is an infinite sequence of global states, Given a context $\gamma$ and a protocol $P$ designed to run in $\gamma$, there is a unique set $\mathcal{R} = \mathcal{R}(P, \gamma)$, of all possible runs of $P$ in $\gamma$. This set is called a *system*, and we study how knowledge evolves in systems. The reason why it does not suffice to consider just one system—say the system consisting of all possible runs in $\gamma$—is because the protocol being executed plays an important role in determining what is known. Typically, the information inherent in receiving a particular message (or in not receiving one) depends on the protocol being used.

---
[3]Depending on the application, a context can include additional components. See [11] for proper exposition.

The essential elements of the model are the following.

- We assume that agents can receive external inputs from the outside world. These are determined in a genuinely nondeterministic fashion, and are not correlated with anything that comes before in the execution or with external inputs of other agents.

- The set of agents is denoted by $\mathbb{P}$. The network consists of the weighted channels graph $\mathsf{Net} = (\mathbb{P}, \mathbb{C}, b)$ in which the weight of a channel $(i,j) \in \mathbb{C}$ consists of a discrete upper bound $b_{ij} \geq 1$. A copy of the network, as well as the current global time, are part of every agent's local state at all times.

- The scheduler, which we typically call the *environment*, is in charge of choosing these external inputs, and of determining message transmission times. The latter are also determined in a nondeterministic fashion, subject to the constraint that delivery satisfies the transmission bounds $b_{ij}$, and messages take at least one time step to be delivered.

- Time is identified with the natural numbers, and agents are assumed to take steps only at integer times. For simplicity, the agents follow deterministic protocols. Hence, a given protocol $P$ for the agents and a given behavior of the environment completely determine the run.

- Events are sends, receives, arrivals of external inputs, and internal actions. All events in a run are distinct, and we denote a generic event by the letter $e$. For ease of exposition, we will assume that an agent's local state contains the set of response actions that the agent has performed. This assumption is needed only for the analysis of response problems, and can be obtained by adding an auxiliary variable keeping track of the history, to each of the agents.

We denote a context satisfying the above assumptions by $\gamma^{\mathsf{max}}$, and use $\mathcal{R}^{\mathsf{max}}$ to denote a system $\mathcal{R}(P, \gamma^{\mathsf{max}})$ consisting of the set of all runs of some protocol $P$ in synchronous context $\gamma^{\mathsf{max}}$.[4]

Note that our model requires transmission times to obey the bounds specified in $\mathsf{Net} = (\mathbb{P}, \mathbb{C}, b)$, but it does not require the agents to have access to a global clock, or to any clocks at all. Nevertheless, the results will apply even in the case in which agents do share a precise global clock, and each agent is scheduled to move at every time step.

---

[4] We defer the rather tedious technical definition of $\gamma^{\mathsf{max}}$ for the full paper.

## 2.2 Syncausality and time bound guarantees

Messages and message chains are a primary tool in coordinating actions in a distributed system. In synchronous networks, in addition to messages, silence can also be used to transmit information. Indeed, as suggested by Lamport in [16], in the synchronous context $\gamma^{\mathsf{max}}$, it is possible to consider the fact that a agent $i$ does *not* send a message over the channel $(i,j) \in \mathbb{C}$ at time $t$ as the sending of a *null* message over the channel. This null message is "received" by $j$ at time $t + b_{ij}$. Motivated by this idea, we proposed the notion of *syncausality* (in [3]), generalizing Lamport's happened-before relation ([15]) to capture generalized message chains consisting of actual messages and null message. Since "not receiving at $t + b_{ij}$" is not an explicit event, it is convenient to define the syncausality relation between agent-time nodes rather than between events. An *agent-time node* (or simply *node*) is a pair $\theta = \langle i, t \rangle$, where $i$ is a agent and $t$ is a time. Such a node represents the instant at time $t$ on $i$'s timeline. Formally, syncausality is defined as follows:

DEFINITION 1 (SYNCAUSALITY). *The **Syncausality** relation in a given run $r$ is the smallest relation $\leadsto_r$ satisfying:*

**Locality:** *If $t \leq t'$ then $\langle i, t \rangle \leadsto_r \langle i, t' \rangle$;*

**Send-rcv:** *If a message sent at $\langle i, t \rangle$ is received at $\langle j, t' \rangle$ then $\langle i, t \rangle \leadsto_r \langle j, t' \rangle$;*

**Null msg:** *If no message is sent over $(i,j) \in \mathbb{C}$ at time $t$ then $\langle i, t \rangle \leadsto_r \langle j, t + b_{ij} \rangle$; and*

**Transitivity:** *If $\theta \leadsto_r \theta'$ and $\theta' \leadsto_r \theta''$, then $\theta \leadsto_r \theta''$.*

Syncausality captures a notion of direct information flow via (generalized) message chains. If $\langle i, t \rangle \not\leadsto_r \langle j, t' \rangle$, then $j$ at time $t'$ does not have information regarding which nondeterministic (or spontaneous) events occur at $\langle i, t \rangle$. A straightforward but useful property of $\leadsto_r$ is:

FACT 1. *If $\langle i, t \rangle \neq \langle j, t' \rangle$ and $\langle i, t \rangle \leadsto_r \langle j, t' \rangle$, then $t < t'$.*

The second (Send-rcv) clause of the definition makes syncausality run-dependent, as actual delivery times depend on the adversary's actions. Hence the subscript $r$ in the $\leadsto_r$ symbol. While syncausality captures direct information flow, the upper bounds on message transmission times allow agents to know about events at remote sites in a less direct fashion. Namely, if $h$ knows about a message sent by $i$ at time $t$ to $j$, then after sufficient time has passed $h$ can be guaranteed that $j$ received $i$'s message. Moreover, if the protocol specifies that $j$ will perform particular actions after receiving this message, then $h$ can know about actions of $j$ without direct information flow from $j$. This can enable them to coordinate their actions without communicating directly. The interaction between communication and time is based on a combination of syncausality and the *bound guarantee* relation, a second causal relation between agent-time nodes that is based on time bounds. Denote by $\delta(i,j)$ the shortest distance between $i$ and $j$ in the weighted graph $\mathsf{Net}$. Intuitively, if we think of a shortest path from $i$ to $j$ in $\mathsf{Net}$ as an "overlay channel" between $i$ and $j$, then $\delta(i,j)$ would be the upper bound for the transmission time over this channel. We define the *bound guarantee* relation as follows:

DEFINITION 2 (BOUND GUARANTEE [3]). *With respect to a network* $\mathsf{Net} = (\mathbb{P}, \mathbb{C}, b)$, *we write* $\langle i, t \rangle \dashrightarrow \langle j, t' \rangle$ *iff* $t + \delta(i, j) \leq t'$.

Intuitively, $\langle i, t \rangle \dashrightarrow \langle j, t' \rangle$ holds, then a message chain initiated at $\langle i, t \rangle$ can be guaranteed to reach $j$ by $\langle j, t' \rangle$. No explicit acknowledgement from $j$ is needed! Put another way, $\langle j, t' \rangle$ is sure to be within the cone of (causal) influence of events that occur at $\langle i, t \rangle$. While syncausality is sensitive to actually realized transmission times, the bound guarantee relation is not. It depends solely on the weighted network $\mathsf{Net}$. This is one of the reasons why bound guarantees provides cross-site information of a type that is not available, for example, in asynchronous settings. In a precise sense, bound guarantees capture the run-invariant part of syncausality that is based solely on $\mathsf{Net}$:

FACT 2. *If* $\langle i, t \rangle \dashrightarrow \langle j, t' \rangle$ *then* $\langle i, t' \rangle \rightsquigarrow_r \langle j, t' \rangle$, *for every run $r$.*

## 2.3 Definition of knowledge

We focus on a very simple logical language in which the set $\Phi$ of primitive propositions consists of propositions of the form $\mathsf{occ}(e)$ for events $e$ of interest. To obtain the logical language $\mathcal{L}$, we close $\Phi$ under propositional connectives and knowledge formulas. Thus, $\Phi \subset \mathcal{L}$, and if $\varphi \in \mathcal{L}$, $i \in \mathbb{P}$, and $G \subseteq \mathbb{P}$, then $\{K_i \varphi, C_G \varphi\} \subset \mathcal{L}$.[5] The formula $K_i \varphi$ is read *agent $i$ knows $\varphi$*, and $C_G \varphi$ is read *$\varphi$ is common knowledge to $G$*. The truth of formulas is evaluated with respect to a triple $(R, r, t)$ consisting of a system $R$, a run $r \in R$, and a time $t \in \mathbb{N}$, and we use $(R, r, t) \vDash \varphi$ to state that $\varphi$ holds at time $t$ in run $r$, with respect to system $R$. Denoting by $r_i(t)$ agent $i$'s local state at time $t$ in $r$, we inductively define

- $(R, r, t) \vDash \mathsf{occ}(e)$    if the event $e$ occurs in $r$ at a time $t' \leq t$;

- $(R, r, t) \vDash K_i \varphi$    if $(R, r', t) \vDash \varphi$ for every run $r'$ satisfying $r_i(t) = r'_i(t)$;

- $(R, r, t) \vDash C_G \varphi$    if $(R, r, t) \vDash K_{i_h} K_{i_{h-1}} \cdots K_{i_1} \varphi$ holds for every $h > 0$ and every sequence $i_h, i_{h-1}, \ldots, i_1$ of agents in $G$.

Given the system $R$, the local state determines what facts are known. Intuitively, a fact $\varphi$ is common knowledge to $G$ if everyone in $G$ knows $\varphi$, everyone knows that everyone knows $\varphi$, and so on *ad infinitum*. We remark that for singleton sets $G = \{i\}$, the operators $C_{\{i\}}$ and $K_i$ coincide.

## 2.4 Centipedes and Brooms

In [3] we introduced *Ordered Response* (OR) and *Simultaneous Response* (SR), two coordination tasks that were simpler than the OJR and GOR problems studied here. We then uncovered the epistemic states communication structures that they necessitate. An instance $\mathsf{OR}\langle e_\mathsf{s}, \alpha_1, \ldots, \alpha_k \rangle$ of ordered

---
[5]This is a simplified logical language for ease of exposition.

response requires that, following occurrence of the triggering event $e_\mathsf{s}$ the set $\{\alpha_1, \ldots, \alpha_k\}$ of responses be performed in a linear temporal order. A central result of [3] is that every run of a protocol solving ordered response must contain a causal structure called a *centipede*:

DEFINITION 3 (CENTIPEDE). *Let* $r \in \mathcal{R}^{\mathsf{max}}$, *let* $\{i_0, \ldots, i_k\} \subseteq \mathbb{P}$, *and let* $t \leq t'$. *A* **centipede** *for* $\langle i_0, \ldots, i_k \rangle$ *in the interval* $(r, t..t')$ *is a sequence* $\theta_0 \rightsquigarrow_r \theta_1 \rightsquigarrow_r \cdots \rightsquigarrow_r \theta_k$ *of nodes such that* (a) $\theta_0 = \langle i_0, t \rangle$, (b) $\theta_k = \langle i_k, t' \rangle$, *and* (c) $\theta_h \dashrightarrow \langle i_h, t' \rangle$ *holds for* $h = 1, \ldots, k-1$.

A centipede is illustrated in Figure 2. The squiggly arrows depict syncausal (message) chains, while the dashed arrows stand for bound guarantees. In a precise sense, a centipede plays in the synchronous context a role analogous to that of message chains in asynchronous ones. In the asynchronous context, a response to the trigger in a protocol ensuring ordered response can occur only if a message chain from the trigger, passing through all previous responses, arrives at the acting agent. In our synchronous model, if $e_\mathsf{s}$ occurs at $\langle i_0, t \rangle$ and $\alpha_h$ is performed at time $t_h$ in $r$, then there must be a centipede for $\langle i_0, \ldots, i_h \rangle$ in $(r, t..t_h)$.

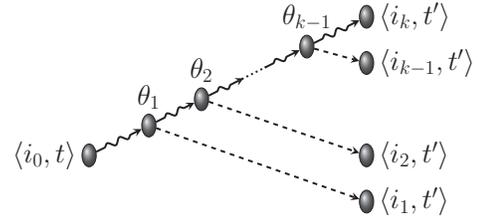

Figure 2: A centipede for $\langle i_0, \ldots, i_k \rangle$ in $(r, t..t')$.

A related causal structure, called a *broom* governs simultaneous coordination. The simultaneous response problem requires all responses to the trigger to occur simultaneously. The responders can act at time $t'$ in response to a trigger at $\langle i_0, t \rangle$ only if a broom structure as in Figure 3 exists.

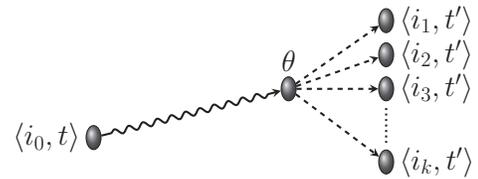

Figure 3: A broom for $\langle i_0, \{i_1 \ldots, i_k\} \rangle$ in $(r, t..t')$.

In a seminal result, Chandy and Misra showed that (Lamport) message chains are a prerequisite for attaining nested knowledge in asynchronous systems [7]. In our synchronous model, centipedes replace message chains in this role:

THEOREM 1 (KNOWLEDGE GAIN, [3]). *Let $P$ be a deterministic protocol, let $r \in \mathcal{R}^{\mathsf{max}} = \mathcal{R}(P, \gamma^{\mathsf{max}})$, and let $e_{\mathsf{s}}$ be an external input received in $r$ at $\langle i_0, t \rangle$. If*

$$(\mathcal{R}^{\mathsf{max}}, r, t') \vDash K_{i_k} K_{i_{k-1}} \cdots K_{i_1} \mathsf{occ}(e_{\mathsf{s}})$$

*then there is a centipede for $\langle i_0, \ldots, i_k \rangle$ in $(r, t..t')$.*

Moreover, the synchronous model goes beyond the asynchronous one by also allowing for common knowledge gain. We have shown that common knowledge gain requires the existence of a broom.

THEOREM 2 (COMMON KNOWLEDGE GAIN, [3]). *Let $P$ be a deterministic protocol, let $r \in \mathcal{R}^{\mathsf{max}} = \mathcal{R}(P, \gamma^{\mathsf{max}})$, let $G \subseteq \mathbb{P}$, and let $e_{\mathsf{s}}$ be an external input received in $r$ at $\langle i_0, t \rangle$. If*

$$(\mathcal{R}^{\mathsf{max}}, r, t') \vDash C_G \mathsf{occ}(e_{\mathsf{s}})$$

*then there is a broom for $\langle i_0, G \rangle$ in $(r, t..t')$.*

## 3. BROOMS & THE DISTRIBUTED SNAPSHOT PROBLEM

Before embarking on the technical analysis of OJR and GOR problems, we now illustrate how a causal analysis (of the "shape" of solutions) can guide the development of efficient solutions to natural problems. We do this by describing the derivation of an optimal solution to the *Synchronous Global Snapshot* problem,[6] a variant of Chandy and Lamport's *Asynchronous Global Snapshot* problem [6]. Due to space constraints, the discussion will be somewhat informal. A formal version appears in [2] and is left for the full paper. A global snapshot of the system at a given time $t$ in a particular run $r$, which we will denote by $\textsc{Snap}(r, t)$, consists of an instantaneous description of the local states of all agents in the system, as well as the contents of the communication channels, at that point in the run. Mechanisms for recording global states come in useful, for example, in association with recovery from system failure. In fact, many applications use such algorithms in order to retain "checkpoints": global states that can be "rolled back" into, when failure occurs (see [20]). Whereas in asynchronous systems the snapshot can only be approximated (see [6]), in systems with a global clock it is possible to compute $\textsc{Snap}(r, t)$ precisely. Indeed, since agents have access to a global clock, if they keep track of their full history, then such a snapshot can be recorded without the need of communication. But the cost of doing this is prohibitive. A natural solution is to record periodical snapshots every $X$ rounds, say. More flexible would be a solution that allows snapshots to be initiated spontaneously, whenever there is good reason to do so. E.g, when some major transaction is completed, or when there is an external indication of an impending storm, requiring a snapshot to be taken.

Consider the problem of taking a spontaneously-generated snapshot. If each agent records its own local state in a global snapshot, then the recording actions are a simultaneous response to the snapshot trigger. By Theorem 2 this requires

---
[6]We are thankful to Gadi Taubenfeld for suggesting this question.

```
OptimalDistributedSnapshot:     % code for agent i
01  Snap_Time_i ← ∞;
02  while True do
03    if time_i = Snap_Time_i then
04      STATE_i ← local state;
05      Snap_Time_i ← ∞;
06    else if ext_Snap or Snap_msg_j(T_j) msgs arrived then
07      candidate_i ← min{T_j : received Snap_msg_j(T_j)};
08      candidate_i ← min{candidate_i, time_i + Rad(i)};
09      if candidate_i < Snap_Time_i then
10        Snap_Time_i ← candidate_i;
11        broadcast Snap_msg_i(Snap_Time_i) to neighbors.
12  end while
```

Figure 4: An Optimal Distributed Snapshot Protocol

a causal broom structure with respect to the arrival of a spontaneous ext_Snap external triggering message. We now describe an optimal distributed snapshot protocol, for the model $\gamma^{\mathsf{max}}$ when agents share a global clock. The code for the protocol appears in Figure 4. For every agent $j \in \mathbb{P}$, define

$$\mathsf{Rad}(j) = \max\{\delta(j, h) : h \in \mathbb{P}\}.$$

Each agent $i$ maintains a local variable named $\mathsf{Snap\_Time}_i$, which is initially set to $\infty$. A time-efficient solution would work as follows: Suppose that a spontaneous snapshot request appears at $(i_0, t_0)$. Then if $t_0 + \mathsf{Rad}(i_0) < \mathsf{Snap\_Time}_{i_0}$, agent $i_0$ sets $\mathsf{Snap\_Time}_{i_0}$ to $t_0 + \mathsf{Rad}(i_0)$ and initiates a flooding of the network by sending a "Snap_msg" labelled with $\mathsf{Snap\_Time}_i$. When agent $i$ receives a snap request labelled by a snap time $T_j$, it compares the current value of $\mathsf{Snap\_Time}_i$ with $t_j + \mathsf{Rad}(j)$ and with $T_j$. If $T_j$ (or $t_i + \mathsf{Rad}(i)$) is smaller than $\mathsf{Snap\_Time}_i$, then agent $i$ updates $\mathsf{Snap\_Time}_i$ to the lower value and initiates a flooding of the network with a $\mathsf{Snap\_msg}(\mathsf{Snap\_Time}_i)$ request. Finally, agents record their local states at the earliest time for which they received a snap message. (In order to account for the contents of the network's channels, they proceed to record messages received on incoming channels until the channels' bounds are met.)

It is easy to see that every agent initiates at most one flooding in this algorithm, though in practice much fewer will be initiated. Moreover, the local states are recorded simultaneously, at the earliest time at which a broom exists for the arrival of external ext_Snap message. This ensures correctness. Finally, a broom is formed in a run of this protocol iff one would be formed in the corresponding run (in the sense that all transmission times and external ext_Snap messages are the same), of a full-information protocol and so this protocol is optimally fast in all cases. No protocol could beat this one, on any run when comparing corresponding runs. Formally, we obtain

THEOREM 3. *The* Optimal Distributed Snapshot *protocol of Figure 4 is **all-case optimal**: For every behavior of nature it records the state as soon as any protocol can.*

## 4. RELATING KNOWLEDGE & ORDERED SIMULTANEOUS RESPONSES

We now define the Ordered Joint Response problem more formally.

DEFINITION 4 (ORDERED JOINT RESPONSE). *Let $e_s$ be an external input and let $A^1, \ldots, A^k$ be disjoint sets of responses. A protocol $P$ solves the instance $\mathsf{OJR}\langle e_s, A^1, \ldots, A^k \rangle$ of ordered joint response if $P$ guarantees for every $h \leq k$ that in every run $r$ in which some response $\alpha \in A^h$ takes place the following conditions are met:*

**Triggering:** *The trigerring event $e_s$ and all responses in $A^1 \cup \cdots \cup A^h$ occur in $r$; and*

**Simultaneity:** *All responses in the same set $A^g$ are performed simultaneously, for $1 \leq g \leq h$; and*

**Linear Ordering:** $t_0 \leq t_1 \leq \cdots \leq t_h$, *where $t_0$ is the time that $e_s$ occurs in $r$ and $t_g$ is the time at which responses in $A^g$ do, for $g = 1, \ldots, h$.*

The simultaneous response problem SR of [3] coincides with a particular subcase of OJR in which $k = 1$: Following the occurrence of the triggering event, all responses must be performed simultaneously. Similarly, the ordered response problem is also a sub-case of OJR, one in which $|A^h| = 1$ for all $h \leq k$.[7] Consider the shape of solutions satisfying $\mathsf{OJR}\langle e_s, A_1, A_2 \rangle$, an instance with $k = 2$. Clearly, for every $\alpha_1 \in A_2$ and $\alpha_2 \in A_2$ occurring in a run $r$ the protocol must solve ordered response and thus produce an appropriate centipede. Moreover, for each of $A_1$ and $A_2$ the protocol must solve simultaneous response, producing a broom. Does a solution need only to produce all of these induced centipedes and the two brooms? We shall show that more is required. Solutions satisfying OJR are associated with a particular shape that combines centipedes and brooms in a natural way. To show this, we apply the connection between simultaneity and common knowledge.

As has been well-established in the literature, simultaneously coordinated actions are intimately connected to common knowledge: When they are performed, the participants have common knowledge of this, and they also have common knowledge that all preconditions of the actions have been satisfied [10, 11, 14]. In the case of OJR, we can show that a particular nested common knowledge formula is a necessary condition for coordinated action. In what follows we denote the set of agents related to the response cluster $A^h$ by $\mathsf{I}^h \subseteq \mathbb{P}$, for all $h \leq k$.

THEOREM 4. *Assume that $P$ is a deterministic protocol solving the instance $\mathsf{OJR}\langle e_s, A^1, \ldots, A^k \rangle$, let $r \in \mathcal{R}^{\mathsf{max}} = \mathcal{R}(P, \gamma^{\mathsf{max}})$, and let $1 \leq h \leq k$. If the responses in $A^h$ occur at time $t_h$ in $r$, then*

$$(\mathcal{R}^{\mathsf{max}}, r, t_h) \vDash C_{\mathsf{I}^h} C_{\mathsf{I}^{h-1}} \cdots C_{\mathsf{I}^1} \mathsf{occ}(e_s).$$

---
[7] Actually OJR is defined along weaker constraints: if the responses occur they must do so simultaneously, whereas in SR and OR the responses must occur in every run where the trigger event $e_s$ occurs.

PROOF. *(Sketch:)* Using the notations in the theorem statement, we prove by induction on $h \geq 1$ that for all $t'_h \geq t_h$: $(\mathcal{R}^{\mathsf{max}}, r, t'_h) \vDash C_{\mathsf{I}^h} C_{\mathsf{I}^{h-1}} \cdots C_{\mathsf{I}^1} \mathsf{occ}(e_s)$.
The results of [10] imply that when an action joint to $\mathsf{I}^h$ is performed, the members of $\mathsf{I}^h$ have common knowledge that it is being performed. The fact that each agent $j \in \mathsf{I}^h$ is assumed to recall the responses it performed (see Section 2) means that this common knowledge is maintained at all times $t'_h > t_h$. The claim follows inductively from the fact that $A^1$ is performed only if $e_s$ occurs, while $A^h$ for $h > 1$ is performed only at or after $A^{h-1}$ has been performed, and so the corresponding subformula for $h - 1$ holds. □

Just as centipedes are closely related to ordered coordination and to nested knowledge formulas, and brooms correspond to common knowledge and simultaneous coordination, a natural composition of the two, which we call a *centibroom*, captures nested common knowledge, and, in turn, linearly ordered clusters of simultaneous actions. Formally,

DEFINITION 5 (CENTIBROOM). *Let $r \in \mathcal{R}^{\mathsf{max}}$, let $\mathsf{I}^h \subseteq \mathbb{P}$ for $1 \leq h \leq k$. A **centibroom** for $\langle i_0, \mathsf{I}^1, \ldots, \mathsf{I}^k \rangle$ in $(r, t..t')$ is a sequence of nodes $\theta_0 \leadsto_r \theta_1 \leadsto_r \cdots \leadsto_r \theta_k$ such that $\theta_0 = \langle i_0, t \rangle$, and $\theta_h \dashrightarrow \langle i_m^h, t' \rangle$ holds for all $h = 1, \ldots, k$ and $i_m^h \in \mathsf{I}^h$.*

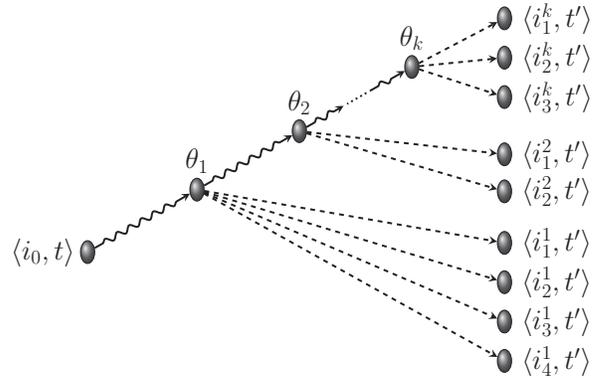

Figure 5: A centibroom for $\langle i_0, \mathsf{I}^1, \ldots, \mathsf{I}^k \rangle$ in $(r, t..t')$.

A centibroom for $\langle i_0, \mathsf{I}^1 \rangle$ is one in which $k = 1$ and there is only one node $\theta_1$. This is a *broom* (see Fig. 1(b)). As mentioned, brooms were shown in [3] to be closely related to common knowledge gain, and to coordinating a Simultaneous Response. A centibroom can be viewed as a generalized centipede, in which every "leg" is replaced by a broom structure.[8]

The Nested Common Knowledge Gain Theorem, our main technical result in this paper, now follows. The theorem shows that, in terms of communication, nested common knowledge requires, at the very least, the existence of a centibroom among the involved agents.

---
[8] The structure we now call a broom was originally called a *centibroom* in [3]. We have since changed the terminology because what is now called a centibroom consists of a centipede whose legs are replaced by brooms.

THEOREM 5 (NESTED COMMON KNOWLEDGE GAIN).
*Let $P$ be a deterministic protocol, $\mathsf{I}^h \subseteq \mathbb{P}$ for $h = 1 \ldots k$, and let $r \in \mathcal{R}^{\mathsf{max}} = \mathcal{R}(P, \gamma^{\mathsf{max}})$. Assume that $e_s$ is the arrival of an external input at $\theta = \langle i_0, t \rangle$ in $r$. If*

$$(\mathcal{R}^{\mathsf{max}}, r, t') \vDash C_{\mathsf{I}^k} C_{\mathsf{I}^{k-1}} \cdots C_{\mathsf{I}^1} \mathsf{occ}(e_s),$$

*then there is a centibroom for $\langle i_0, \mathsf{I}^1, \ldots, \mathsf{I}^k \rangle$ in $(r, t..t')$.*

PROOF. Assume the notations and conditions of the theorem. We will use $\mathsf{I}^h = \{i_1^h, \ldots, i_{s_h}^h\}$ to denote the set of agents $\{i | \langle i, a \rangle \in A^h\}$ participating in the responses $A^h$, for every $h < k$. Let $\bar{\mathcal{K}}_{\mathsf{I}^h} = K_{i_1^h} \cdots K_{i_{s_h}^h}$ denote the string of (nested) knowledge operators spanning the agents in $\mathsf{I}^h$ in sequence. We write $(\bar{\mathcal{K}}_{\mathsf{I}^h})^m$ to denote $m$ consecutive copies of $\bar{\mathcal{K}}_{\mathsf{I}^h}$. Denote $d = t' - t + 1$. The fact that $(\mathcal{R}^{\mathsf{max}}, r, t') \vDash C_{\mathsf{I}^k} C_{\mathsf{I}^{k-1}} \cdots C_{\mathsf{I}^1} \mathsf{occ}(e_s)$ holds implies by definition of common knowledge that

$$(\mathcal{R}^{\mathsf{max}}, r, t') \vDash (\bar{\mathcal{K}}_{\mathsf{I}^k})^d \cdot (\bar{\mathcal{K}}_{\mathsf{I}^{k-1}})^d \cdots (\bar{\mathcal{K}}_{\mathsf{I}^1})^d \mathsf{occ}(e_s).$$

By Theorem 1 (Knowledge Gain), there is a centipede

$$\begin{aligned}
\sigma = & \langle i_0, t \rangle \leadsto_r \\
& \theta_1^k \leadsto_r \cdots \leadsto_r \theta_{d \cdot s_k}^k \leadsto_r \\
& \theta_1^{k-1} \leadsto_r \cdots \leadsto_r \theta_{d \cdot s_k}^k \leadsto_r \\
& \cdots \\
& \theta_1^1 \leadsto_r \cdots \leadsto_r \theta_{d \cdot s_1}^1
\end{aligned}$$

for $\langle i_0, (i_1^k, \ldots, i_{s_k}^k)^d, \ldots, (i_1^1, \ldots, i_{s_1}^1)^d \rangle$ in $(r, t..t')$.

We partition the centipede $\sigma$ into the segments $\Theta_k$ to $\Theta_1$ such that $\Theta_h = \langle \theta_1^h, .., \theta_{d \cdot s_h}^h \rangle$. Thus, each $\Theta^h$ corresponds to the $\bar{\mathcal{K}}_{\mathsf{I}^h}$ portion of the formula. Note that if $h < k$ then $\theta \leadsto_r \theta'$ for every $\theta \in \Theta_h$ and $\theta' \in \Theta_{h+1}$. Moreover, denoting $\theta = (i_\theta, t_\theta)$ and $\theta' = (i_{\theta'}, t_{\theta'})$, by Fact 1 we obtain that if $\theta \neq \theta'$ then $t_\theta < t_{\theta'}$. It follows that there can be at most $t' - t + 1 = d$ distinct nodes $\beta_1 \leadsto_r \beta_2 \leadsto_r \cdots \leadsto_r \beta_\ell$ in $\sigma$, and in particular at most $d$ distinct nodes in every segment $\Theta_h$ of $\sigma$.

Given $h \leq k$, recall that $s_h = |\mathsf{I}^h|$ and that $\theta_\ell^h \dashrightarrow i_{(\ell \bmod s_h)+1}^h$ holds for each $\ell \leq d \cdot s_h$. As the segment $\Theta_h$ contains $d \cdot s_h$ nodes of which at most $d$ are distinct, by the pigeonhole principle there must exist some node $\beta_h \in \Theta_h$ such that $\beta_h = \theta_x^h = \theta_{x+1}^h = \cdots = \theta_{x+s_h-1}^h$ for some $x \in [1..s_h \cdot d]$. By definition of centipede and the structure of our particular centipede $\sigma$, we get that $\beta_h \dashrightarrow \langle i_{x+\delta (\bmod s_h)}^h, t' \rangle$ for all $\delta \in [0..s_h - 1]$. It thus follows that $\beta_h \dashrightarrow \langle i, t' \rangle$ for all $i \in \mathsf{I}^h$. As noted above, we also have that $\theta_0 \leadsto_r \beta_1 \leadsto_r \beta_2 \leadsto_r \cdots \leadsto_r \beta_k$. We conclude that $\langle \theta_0, \beta_1, .., \beta_k \rangle$ is a centibroom for $\langle i_0, \mathsf{I}^1, \ldots, \mathsf{I}^k \rangle$ in $(r, t..t')$, as desired. □

Theorem 5 presents a strict and significant generalization of both the Knowledge Gain Theorem (Theorem 1 above) and of the Common Knowledge Gain Theorem of [3] to the case of nested *common knowledge*. It is the first nontrivial and useful nested CK gain theorem that we are aware of. Recall from Theorem 4 that nested common knowledge is a prerequisite for action in OJR problems. Combining the two theorems, we obtain a strict generalization of both the Centipede Theorem and the Broom Theorem of [3], matching centibrooms with ordered joint response.

COROLLARY 1 (CENTIBROOM THEOREM). *Assume that $P$ satisfies the $\mathsf{OJR}\langle e_s, A^1, \ldots, A^k \rangle$ property, that $e_s$ occurs at $\langle i_0, t \rangle$ in $r \in \mathcal{R}(P, \gamma^{\mathsf{max}})$. Denote by $\mathsf{I}^m$ the set of agents responding in $A^m$, for $m = 1, \ldots, k$. For every $1 \leq h \leq k$, if the responses in $A^h$ are performed at time $t_h$ in $r$, then there is a centibroom for $\langle i_0, \mathsf{I}^1, \ldots, \mathsf{I}^h \rangle$ in $(r, t..t_h)$.*

## 5. CHARACTERIZING GENERAL ORDERED RESPONSE

We define a *response ordering* to be a finite directed graph $\mathsf{Ro} = (V \langle T, A \rangle, \preceq)$ where the set of nodes $V = T \cup A$ is a disjoint union of the set of (externally initiated) triggering events $T$, and the a set of response actions $A$. Moreover, $\preceq$ is a preorder over $A \cup T$ (a reflexive and transitive binary relation), in which the nodes of $T$ are all initial elements. Thus, $\beta \preceq \tau$ for $\tau \in T$ is possible only if $\beta = \tau$. Responses have the form $\alpha = (\mathsf{a}, i)$, where $\mathsf{a}$ is an action to be performed by agent $i$. We define $\mathsf{base}_\alpha$, the *trigger base* of a response $\alpha \in A$, by

$$\mathsf{base}_\alpha = \{e \in T : e \preceq \alpha\}$$

DEFINITION 6 (GENERAL ORDERED RESPONSE).
*A response ordering $\mathsf{Ro} = (V \langle T, A \rangle, \preceq)$ defines an instance $\mathsf{GR}\langle \mathsf{Ro} \rangle$ of the **General Ordered Response** problem. A protocol $P$ solves $\mathsf{GR}\langle \mathsf{Ro} \rangle$ if it guarantees both*

**Triggering:** *A response $\alpha \in A$ occurs in a run iff all of the events in $\mathsf{base}_\alpha$ occur; and*

**Weak Ordering:** *If $\alpha_1 \preceq \alpha_2$, and in a particular run $\alpha_1$ occurs at time $t_1$ while $\alpha_2$ occurs at $t_2$, then $t_1 \leq t_2$.*

Given the weak ordering clause, nodes on a cycle in the response ordering graph are responses that must be performed simultaneously in every solution to the problem. Characterizing the shape of GOR coordination is done by focusing on the ability of GOR to specify that a collection of disjoint sets of responses (we think of them as *clusters*) will be performed such that all responses in a cluster take place simultaneously. Moreover, any linearly ordered set of such clusters, together with an initial triggering event in their base, define an instance of OJR as a subproblem of the given GOR. By combining the above intuition with Theorems 4 and 5 that relate to the OJR problem, we can characterize the causal requirements for general response problems.

When the response ordering $\mathsf{Ro}$ is a DAG, it specifies a *partial order* on the individual responses. Otherwise, it can be viewed as a directed graph, and every directed graph can be decomposed into its strongly connected components (SCCs) [9]. This decomposition naturally induces a graph on the SCCs, which is itself a DAG. Given an instance $\mathsf{GR}\langle A, T, \mathsf{Ro} \rangle$, let $\mathsf{S} = \{\mathsf{scc}_1, \ldots, \mathsf{scc}_k\}$ be the set of strongly connected components of $A \in \mathsf{Ro}$, and let $\mathsf{l}$ be a node labelling such that $\mathsf{l}(\mathsf{scc}_h) = \mathsf{l}^h$ is the set of agents performing responses in $\mathsf{scc}_h$. (In particular, if $\mathsf{scc}_h$ is a single response, then $\mathsf{l}(\mathsf{scc}_h)$ is the agent performing it.) Then

CRO $= \langle S, T, \preceq' \rangle$ where $scc_i \preceq' scc_j$ iff $\alpha_i \preceq \alpha_j$ for some $\alpha_i \in scc_i$ and $\alpha_j \in scc_j$, which we call the *DAG decomposition of* RO.

Going back to Example 2, a part of the detailed response ordering, containing cycles for agent groups that must react simultaneously, is shown in Figure 6. Figure 1, shown earlier, is the SCC decomposition of this full RO.

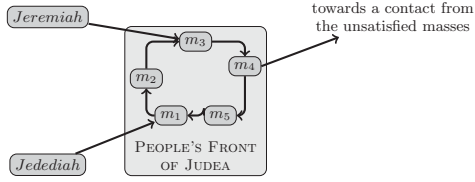

Figure 6: Part of the detailed RO for Judea, 71AD

Formally, we show:

THEOREM 6. *Fix a GOR instance* $\Gamma = GR\langle A, T, RO \rangle$ *and a response* $\alpha \in A$. *Assume that P solves* $\Gamma$, *and let* $r \in \mathcal{R}^{\max} = \mathcal{R}(P, \gamma^{\max})$ *be a run in which $\alpha$ takes place at the time $t'$. Then*

a) *If* RO *is a DAG, then **there exists a centipede** for*
$$\langle i_0, I(\alpha_1), \ldots, I(\alpha_k) \rangle$$
*in $(r, t..t')$ **for every path** $e_0 \preceq \alpha_1 \preceq \alpha_2 \preceq \cdots \preceq \alpha_k = \alpha$ in* RO *such that $e_0$ occurs at the $\langle i_0, t \rangle$ in $r$. More generally,*

b) *In case* RO *is not a DAG, let* CRO $= \langle S, T, \preceq' \rangle$ *be the DAG decomposition of* RO. *Then **there exists a centibroom** for $\langle i_0, I(scc_1), \ldots, I(scc_k) \rangle$ in $(r, t..t')$, **for every path** $e_0 \preceq' scc_1 \preceq' scc_2 \preceq' \cdots \preceq' scc_k = scc_\alpha$ in* CRO *such that $\alpha \in scc_\alpha$ and $e_0$ occurs at $\langle i_0, t \rangle$ in $r$.*

PROOF. *(Sketch:)* Part (a) is an instance of part (b) in which all SCCs are singletons, since a centipede is a centibroom in which every broom contains a single target node. It thus suffices to show part (b). Under the conditions and the notation of the theorem statement, the existence of the path $e_0 \preceq' scc_1 \preceq' scc_2 \preceq' \cdots \preceq' scc_k = scc_\alpha$ in CRO ensures that the protocol $P$ must satisfy the $OJR(e_0, scc_1, \ldots, scc_k)$ property. Denoting $I^h = I(scc_h)$ for $1 \le h \le k$, Theorem 4 implies that $(\mathcal{R}^{\max}, r, t') \vDash C_{I^k} C_{I^{k-1}} \cdots C_{I^1} occ(e_0)$. The claim now follows immediately from Theorem 5 (nested common knowledge gain). □

Notice that a centipede contains a linear chain of syncausally-related agents that mimic the linear temporal ordering that is required of the responses in an Ordered Response. The shape of the OR problem and the "shape" of its solution are closely related. Theorem 6 shows that for more general specifications such as a partial-order GOR, the shapes are not as tightly connected. The partial order implies a set of linear orderings, and the centipedes for these must be constructed. In a precise sense, it is the required logical structure, specified in terms of a conjunction of nested knowledge and nested common knowledge formulas, that constrains the shape of the solution.

Theorem 6 states *necessary* conditions for any solution to GOR problems in a context in which the environment can deliver message subject to given upper bounds on transmission times. In a precise sense, this theorem cannot be strengthened: In the full paper we show (using the same technique as in [4]) that the condition in Theorem 6 is in a precise sense *sufficient*, as well as necessary: In a context in which the agents have access to a global clock there is a full-information protocol solving the GOR, in which each response $\alpha$ is performed at the first time $t'$ at which all the required centibrooms for $\alpha$ by Theorem 6 exist. For every behavior of nature, the resulting protocols ensures that the agents respond in the fastest possible manner.

Using Theorem 6 to analyze Example 2, we obtain that in order for a rebellion to start in Judea we need that all three instigators revolt, and that there exist centibroom communication patterns for each of the following chains of population groups (using *masses* for the unsatisfied masses and *old regime* for supporters of the old regime):

- *Jeremiah* $\preceq'$ *PFJ* $\preceq'$ *masses* $\preceq'$ *old regime*,
- *Jedediah* $\preceq'$ *PFJ* $\preceq'$ *masses* $\preceq'$ *old regime*, and
- *Brian* $\preceq'$ *JPF* $\preceq'$ *masses* $\preceq'$ *old regime*.

We remark that a GOR specifying a partial order on responses (with no simultaneous actions required), as in Theorem 6(a), is implementable in the asynchronous model as well. In the asynchronous model, centipedes reduce to message chains ([3]), and so the message chains must closely follow the paths in the graph of RO in this case (cf. Parikh and Krasucki [21]). This is no longer the case in the synchronous model, since there the centipedes impose a richer and more flexible structure in the shape of GOR implementations.

## 6. CONCLUSIONS

In summary, this paper uses an epistemic analysis to significantly extend our understanding of the interaction between knowledge, time and causality in multi-agent systems. This new understanding can be applied to a broad class of coordination problems, yielding insights and guidance regarding how to design efficient, even optimal, solutions to coordination tasks. Natural extensions currently being explored consider the analysis of coordination tasks stated in terms of explicit time bounds, rather than orderings. For example, if we specify that response $\alpha_2$ must occur no later than 5 days after response $\alpha_1$ has occurred, or even that responses $\alpha_1$ and $\alpha_2$ must occur exactly 3 days apart from each other. These issues are explored in [5] and [12].

The subject of network dynamics, the diffusion of ideas and actions through a social network, has been extensively studied since the seventies, and in particular in the last decade. We believe that our results pertaining to minimal networks will be of value in this ongoing effort. It is interesting to note that the minimal communication graph needed for achieving nested common knowledge in our system is significantly sparser than that needed by Chwe in [8]. Currently our problem formulations are too far apart from Chwe's to allow for rigorous comparison, so further research is needed in order to get to the bottom of this.

The current paper extends and generalizes our previous work in [3], from the study of sequential or strictly simultaneous coordination to GOR problems. The latter allow coordination specified by an arbitrary partial order, or in terms of a partial order defined of clusters, where each cluster of actions is necessarily simultaneous. GOR problems also allow multiple triggering events, and responses must be performed if all of the spontaneous triggering events that they depend on occur. Thus, the dependence on triggers is conjunctive. A natural question that arises from the current investigation is, what would be the effect on required communication if the dependence on triggers could be more general, say defined by a general boolean function. Similarly, perhaps the interdependence among responses could also be specified more generally. Indeed, GOR problems can be specified by a suitably expressive temporal logic [18]. Is there a sensible way of relating general temporal-epistemic formulas to the causal structure required to attain them?

We illustrated the applicability of a causal analysis in terms of syncausality and bound guarantees in Section 3, showing how it can be used to derive an optimal solution to the synchronous distributed snapshot problem. Our new results can similarly allow the synthesis of efficient, even optimal, solutions to many other distributed tasks in synchronous settings. A promising direction for further study is to explore the epistemic underpinnings of particular tasks, and apply a causal analysis in this style, in order to improve the analysis and solutions for such tasks. There is much room for further investigation.

## 7. REFERENCES

[1] C. Attiya and J. Welch. *Distributed Computing: Fundamentals, Simulations, and Advanced Topics, 2nd Edition*. Wiley, 2004.
[2] I. Ben-Zvi. *Causality, Knowledge and Coordination in Distributed Systems*. PhD thesis, Technion, Israel, 2011.
[3] I. Ben-Zvi and Y. Moses. Beyond Lamport's happened-before: On the role of time bounds in synchronous systems. In *DISC 2010*, pages 421–436, 2010.
[4] I. Ben-Zvi and Y. Moses. On interactive knowledge with bounded communication. *Journal of Applied Non-Classical Logics*, 21(3-4):323–354, 2011.
[5] I. Ben-Zvi and Y. Moses. Agent-time epistemics and coordination. In *Proceedings of the 5th Indian conference on logic and its applications*, ICLA'13, 2013.
[6] K. M. Chandy and L. Lamport. Distributed snapshots: determining global states of distributed systems. *ACM Trans. on Computer Systems*, 3(1):63–75, 1985.
[7] K. M. Chandy and J. Misra. How processes learn. *Distributed Computing*, 1(1):40–52, 1986.
[8] M. S.-Y. Chwe. Communication and coordination in social networks. *Review of Economic Studies*, 67(1):1–16, 2000.
[9] T. H. Cormen, C. E. Leiserson, R. L. Rivest, and C. Stein. *Introduction to Algorithms, 3rd ed.* MIT Press, 2009.
[10] R. Fagin, J. Y. Halpern, Y. Moses, and M. Y. Vardi. Common knowledge revisited. *Annals of Pure and Applied Logic*, 96(1-3):89 – 105, 1999.
[11] R. Fagin, J. Y. Halpern, Y. Moses, and M. Y. Vardi. *Reasoning about Knowledge*. MIT Press, 2003.
[12] Y. Gonczarowski and Y. Moses. Timely common knowledge: Characterising asymmetric distributed coordination via vectorial fixed points. In *Proceedings of the 14th Conference on Theoretical Aspects of Rationality and Knowledge*, TARK XIV, 2013.
[13] M. Granovetter. The Strength of Weak Ties. *The American Journal of Sociology*, 78(6):1360–1380, 1973.
[14] J. Y. Halpern and Y. Moses. Knowledge and common knowledge in a distributed environment. *Journal of the ACM*, 37(3):549–587, 1990. A preliminary version appeared in *Proc. 3rd ACM Symposium on Principles of Distributed Computing*, 1984.
[15] L. Lamport. Time, clocks, and the ordering of events in a distributed system. *Communications of the ACM*, 21(7):558–565, 1978.
[16] L. Lamport. Using time instead of timeout for fault-tolerant distributed systems. *ACM Trans. Program. Lang. Syst.*, 6(2):254–280, 1984.
[17] N. A. Lynch. *Distributed Algorithms*. Morgan Kaufmann, 1996.
[18] Z. Manna and A. Pnueli. *The Temporal Logic of Reactive and Concurrent Systems*, volume 1. Springer-Verlag, Berlin/New York, 1992.
[19] S. Morris. Contagion. *Review of Economic Studies*, 67(1):57–78, January 2000.
[20] N. Neves and W. K. Fuchs. Using time to improve the performance of coordinated checkpointing. In *Proceedings of the 2nd International Computer Performance and Dependability Symposium (IPDS '96)*, 1996.
[21] R. Parikh and P. Krasucki. Levels of knowledge in distributed computing. *Sādhanā*, 17(1):167–191, 1992.
[22] T. Schelling. Dynamic models of segregation. *Journal of Mathematical Sociology*, 1(1):143–186, 1971.